\documentclass[lettersize,journal]{IEEEtran}
\usepackage[switch]{lineno}
\usepackage[x11names]{xcolor}
\usepackage[colorlinks,linkcolor=blue,allcolors=blue]{hyperref}
\AtBeginDocument{\hypersetup{citecolor=DodgerBlue4}}

\usepackage{amsmath,amsfonts}
\usepackage{algorithmic}
\usepackage{algorithm}
\usepackage{array}
\usepackage[caption=false,font=normalsize,labelfont=sf,textfont=sf]{subfig}
\usepackage{textcomp}
\usepackage{listings}
\usepackage{xcolor}
\usepackage{verbatim}
\usepackage{fancyvrb} 
\usepackage{amsmath}
\usepackage{amssymb}
\usepackage{gensymb}
\usepackage{amstext}
\usepackage{mathtools}
\usepackage{xcolor}
\usepackage{soul}
\usepackage{esint}
\usepackage{caption}
\usepackage{wrapfig}
\usepackage{float}
\usepackage{dblfloatfix}
\usepackage[symbol]{footmisc}
\modulolinenumbers[1]
\usepackage{geometry}
\usepackage{upgreek}
\usepackage{epsfig}
\usepackage{subcaption}

\usepackage{array}
\usepackage{xcolor}
\usepackage{multirow} 
\usepackage{rotating}
\usepackage{amsfonts}
\usepackage{booktabs}
\usepackage{siunitx}
\usepackage{chemformula}
\newcolumntype{P}[1]{>{\centering\arraybackslash}p{#1}}
\usepackage{amsmath}
\usepackage{makecell}
\usepackage{graphicx} 
\usepackage{url}
\usepackage{verbatim}
\usepackage{graphicx}
\usepackage{cite}
\begin{document}

\title{Integration of Silica in G4CMP for Phonon Simulations: Framework and Tools for Material Integration }
 \author{Caitlyn Stone-Whitehead$^1$, Israel Hernandez$^{2,3}$, Connor Bray$^1$, Allison Davenport$^1$, Spencer Fretwell$^1$, Abigail Gillespie$^1$, Joren Husic$^1$, Mingyu Li$^4$, Andrew Marino$^1$, Kyle Leach$^{1,5}$, Bismah Rizwan$^1$, Wouter Van De Pontseele$^1$, Grace Wagner$^1$\\
Colorado School of Mines, Department of Physics, CO 80401, USA$^1$ \\
Illinois Institute of Technology, Department of Physics, Chicago, IL 60616, USA$^2$\\
Fermi National Accelerator Laboratory, Batavia, IL 60510, USA$^3$\\
Massachusetts Institute of Technology, Department of Physics, Cambridge, MA 02139, USA$^4$\\
Facility for Rare Isotope Beams, Michigan State University, 640 S Shaw Lane, East Lansing, 48824, MI, USA$^5$
\thanks{This paper was produced by the G4CMP Consortium.}
\thanks{Manuscript received June --, 2025; revised -------, 2025.}}

\markboth{IEEE TRANSACTIONS ON APPLIED SUPERCONDUCTIVITY,~Vol.~14, No.~8, August~2021}%
{Shell \MakeLowercase{\textit{et al.}}: A Sample Article Using IEEEtran.cls for IEEE Journals}

\IEEEpubid{0000--0000/00\$00.00~\copyright~2025 IEEE}

\maketitle
\begin{abstract}
Superconducting detectors with sub-eV energy resolution have demonstrated success setting limits on Beyond the Standard Model (BSM) physics due to their unique sensitivity to low-energy events. G4CMP, a Geant4-based extension for condensed matter physics, provides a comprehensive toolkit for modeling phonon and charge dynamics in cryogenic materials. This paper introduces a technical formalism to support the superconducting qubit and low-threshold detector community in implementing phonon simulations in custom materials into the G4CMP. As a case study, we present the results of a detailed analysis of silica phonon transport properties relevant for simulating substrate backgrounds in Beryllium Electron capture in Superconducting Tunnel junctions (BeEST)-style experiments using G4CMP. Additionally, Python-based tools were developed to aid users in implementing their own materials and are available on the G4CMP repository. 
\end{abstract}


\section{Introduction}
\IEEEPARstart{G}{eant4} 
\cite{Geant4,Allison2006,Allison2016}, known as the reference simulation engine for high-energy physics, is a C++ toolkit for developing Monte Carlo simulations of particle interactions with matter. In the past decade, Geant4 has been expanded to the low-energy regime for modeling radiation interactions in microelectronic devices as well as biological media. To leverage this framework for phonon and charge excitation in semiconductor crystals, Geant4 Condensed Matter Physics, G4CMP \cite{G4CMP} was developed to model experimental results from the Cryogenic Dark Matter Search (CDMS) \cite{SuperCDMs}. The G4CMP framework has demonstrated success in reproducing both theoretical and experimental results in cryogenic semiconductor crystals coupled to superconducting devices. The toolkit has continued to expand for a wider range of applications via the inclusion of novel materials, demonstrating usage in optimizing phonon collection and mitigating quasiparticle poisoning effects in qubits \cite{G4CMP2025}. 

To support expanded usage of G4CMP by the superconducting detector community, the technical details of calculating phonon parameters for simulations in G4CMP are described in the sections below. The process for implementing silicon dioxide (both amorphous and $\alpha$-Quartz) is used as an example, motivated by the Beryllium Electron capture in Superconducting Tunnel junctions (BeEST) experiment \cite{Kim2024}. Silica was implemented in G4CMP \newpage \noindent to model the time‑dependent propagation of phonons across the substrate’s thermal oxide layer to reduce systematic uncertainties in the reconstruction of nuclear recoil spectra. Additionally, a tutorial for calculating the phonon density of states (DOS) with mode-dependent contributions has been developed to aid in the repeatability of this work. 

\section{Elastic Constants and Mode Dependent Speed of Sound}
Elastic constants are material-dependent parameters that describe the stress-strain relationship in a given medium. They include the second-order elastic constants (SOECs), which represent the linear relationship between stress and strain, and the third-order elastic constants (TOECs), which represent the nonlinear relationship between them. SOECs and TOECs are of key importance for computing Lamé parameters for anharmonic downconversion as well as reproducing phonon caustics. These values are commonly obtained experimentally from measurements of bulk materials at room temperature \cite{Tamura1985}. For our modeling of silicon dioxide, the SOECs and TOECs are included in \cite{Bogardus1965} and \cite{alphaTOEC}. The mode-dependent speeds of sound are implemented from experimental results using bulk materials \cite{alphaSound,amorphTOEC}. 

\section{Lamé Parameters}
The Lamé parameters offer a parameterization of the SOECs (\(\mu\), \(\lambda\)). We have developed a general formalism to express the Lamé parameters as functions of the SOECs and TOECs (\(\mu\), \(\lambda\), \(\alpha\), \(\beta\), \(\gamma\)). By subsequently applying the symmetries present in these elastic constants \cite{Hearmon}, we derive the expressions for the Lamé parameters, which are specific to $\alpha$-Quartz.
We follow the same formalism used by Fedorov\cite{Fedorov1968} to derive the Lamé parameters as a function of the SOECs and TOECs. We define the following scalar quantity,
\begin{equation}
F=\text{min}(C_{ijkl}-C_{ijkl}^{0})^{2}
\end{equation}
where $C_{ijkl}$ are the measured values of the SOECs and $C_{ijkl}^{0}$ is the isotropic tensor. The isotropic tensor can be written as: 
\begin{equation}
C_{iklm}^{0}=\lambda\delta_{ik}\delta_{lm}+\mu(\delta_{il}\delta_{km}+\delta_{im}\delta_{kl}).
\end{equation}
In the isotropic approximation, the value of \textit{F} must have a minimum. Considering this, then $\partial_{\lambda} F=\partial_{\mu} F=0$. Taking the partial derivative with respect to $\lambda$ and $\mu$, we can then solve the resulting system of equations to obtain the values as a function of elastic moduli:
\begin{equation}
\begin{split}
\mu & =(3C_{lklk}-C_{llkk})/30\\[3mm]
 \lambda & =(2C_{lklk}-C_{llkk})/15.\\
 \end{split} 
\end{equation}
For $\alpha$, $\beta$, and $\gamma$, the same procedure is followed. We consider the following scalar quantity 
\begin{equation}
F=\text{min}(C_{ijklmn}-C_{ijklmn}^{0})^{2}
\end{equation}
where $C_{ijklmn}^{0}$ is the isotropic sixth-rank tensor that can be redefined as $C_{ijklmn}^{0}=A\alpha+B\beta+C\gamma$ \footnote{The expressions for A, B and C can be found here \cite{Campbell_Deem_2020_Lame}. This B is different from equation 10.}. We then square Equation 4:
\begin{multline}
(C_{ijklmn}^{0})^{2}=A^{2}\alpha^{2}+B^{2}\beta^{2}+C^{2}\gamma^{2}\\
          +2\alpha\beta AB+2\alpha\gamma AC+2\gamma\beta BC.
\end{multline}
Calculating the partial derivatives respect to $\alpha$, $\beta$, and $\gamma$, we have the following system of three equations:
\begin{equation}
\begin{split}
 C_{iikknn}=\frac{\partial C_{ijklmn}^{0}}{\partial\alpha}&=2A^{2}\alpha+2AB\beta+2AC\gamma \\
 6C_{iilnlm}=\frac{\partial C_{ijklmn}^{0}}{\partial\beta}&=2B^{2}\beta+2AB\alpha+2BC\gamma \\
 8C_{inilln}=\frac{\partial C_{ijklmn}^{0}}{\partial\gamma}&=2C^{2}\gamma+2AC\alpha+2BC\beta. \\
\end{split}
\end{equation}
Arranging the above system of equations in matrix form and finding the inverse, we have that $\alpha, \beta$, and $\gamma$, can be written as 
\begin{equation}
\begin{split}
\label{eq:Lame_alpha}
\alpha&=\left(8C_{iillnn}-15C_{iilnln}+8C_{inilln}\right)/105 \\[3mm]
\beta&=\left(-5C_{iillnn}+19C_{iilnln}-12C_{inilln}\right)/210\\[3mm]
\gamma&=\left(2C_{iillnn}-9C_{iilnln}+9C_{inilln}\right)/210.\\[3mm]
\end{split}
\end{equation}
$\alpha$-Quartz belongs to the trigonal space group P3221, which contains 7 independent SOECs and 14 independent TOECs.
\begin{table*}[t]
\caption{Anharmonic downconversion rate and additional values needed for implementation in G4CMP. The Lamé parameters ($\mu$, $\lambda$) and TOEC coefficients ($\alpha$, $\beta$, $\gamma$) (units of pressure) are used to compute the anharmonic decay rate coefficient $A$.  We quote a literature value $A_{c,l}$ for comparison to $A$. The $\alpha$-Quartz rows correspond to calculated values for the two sets of elastic constants \cite{alphaTOEC}.}
\centering
\begin{tabular}{m{1cm}m{0.9cm}m{0.9cm}m{1.9cm}m{1.5cm}m{1.9cm}m{0.6cm}m{1.5cm}m{0.9cm}}
\toprule\toprule
Material 
  & $\mu$ [GPa] & $\lambda$ [GPa] & $\alpha$ [GPa] 
  & $\beta$ [GPa] & $\gamma$ [GPa] & $F_{TT}$ 
  & $A$ [$10^{-55}\,\mathrm{s}^4$] 
  & $A_{c,l}$ [$10^{-55}\,\mathrm{s}^4$] \\
\midrule
\ch{Amorphous} 
  & 31.2 \cite{Tamura1985} & 16.1 \cite{Tamura1985} 
  & 428.42 \cite{Tamura1985} & 54.18 \cite{Tamura1985} 
  & –13.14 \cite{Tamura1985} & 0.53 
  & 8.23 & 7.99 \cite{Tamura1985} \\

\ch{$\alpha$-Quartz} 
  & 45.9 & 3.97 & –155.59 & 15.64 & –97.99 & 0.71 
  & 36.69 & $\cdot$ \\

\ch{$\alpha$-Quartz} 
  & 48.7 & 5.47 & –197.32 $\pm$ 2.97 & 51.28 $\pm$ 3.83 
  & –138.92 $\pm$ 1.73 & 0.72 
  & 19.28 $\pm$ 0.08 & $\cdot$ \\
\bottomrule\bottomrule
\end{tabular}

\vspace{1em}
\label{table:Anharmonic_and_Isotopic}
\end{table*}

Using the symmetry group to reduce the expression for the Lamé constants, the resulting equations are:
\begin{equation}
\begin{split}
\lambda &= \frac{1}{15} ( 2C_{11} + 4C_{12} + 8C_{13} \\
&-4C_{44} + C_{33} - 2C_{66}) \\[3mm]
\mu &= \frac{1}{15} ( 2C_{11} - C_{12} - 2C_{13} \\
&+ C_{33} + 6C_{44} + 3C_{66} )
\end{split}
\end{equation}
\begin{equation}
\begin{split}
&\alpha =  (-17C_{111} + 30C_{112} +33C_{113} + 78C_{123} \\ 
& -57C_{133} - 156C_{144} + 84C_{155}\\
&-17C_{222} + C_{333} - 12C_{344})/105 \\[3mm]
&\beta = (71C_{111} + 4C_{112} - 11C_{113} - 68C_{123} \\ 
& + 103C_{133} + 220C_{144}- 140C_{155} \\
&- 55C_{222} + 2C_{333} + 4C_{344})/210 \\[3mm]
&\gamma =(-34C_{111} - 24C_{112} +3C_{113} + 30C_{123} \\
& - 51C_{133}- 144C_{144}+126C_{155}\\
&+ 92C_{222} + 2C_{333} + 18C_{344})/210. \\
\end{split}
\end{equation}

The calculated Lamé parameters for $\alpha$-Quartz and amorphous silicon dioxide are shown in Table \ref{table:Anharmonic_and_Isotopic}. The results for amorphous silicon dioxide are calculated from the isotropic expressions for the Lamé constants \cite{Tamura1985}. 

\newpage
\section{Anharmonic Downconversion}
 Anharmonic downconversion refers to the redistribution of a longitudinal phonon's energy into two lower energy phonons due to deviations from the quadratic form of the interatomic potential. The initial longitudinal phonon's energy can be redistributed into two transverse phonons or into a longitudinal and a transverse phonon. The decay rates for both decay modes are required to model this process. Once the Lamé parameters have been reduced for the given symmetries in a crystal space group, the anharmonic downconversion rates can be calculated using the Tamura formalism \cite{Tamura1985}. A Python program has been developed for this purpose that only requires the Lamé parameters (GPa), speeds of sound (m/s), and density ($kg/m^3$). The code is will soon be included in the G4CMP GitHub repository under tools. The results of this computation for silicon dioxide are shown in Table \ref{table:Anharmonic_and_Isotopic}.

\section{Isotopic Scattering}
Isotopic scattering refers to the scattering of phonons from naturally occurring isotopic impurities in a crystal. As phonons propagate within a crystalline solid, deviations from isotopic mass variations serve as scattering centers. Isotopic disorder introduces local mass fluctuations that disrupt the periodicity of the lattice, thereby enabling phonon scattering. Treating these mass fluctuations as a perturbation to the ideal harmonic lattice Hamiltonian, one can employ time-dependent perturbation theory to derive the phonon scattering rate associated with this mechanism \cite{srivastava1990physics}. The resulting scattering rate exhibits a quartic dependence on phonon energy, reflecting the strong energy sensitivity of this process.
\begin{align}
\label{eq:IsotopicScattering1}
\Gamma_{\mathrm{isotopic}} =\ & B\nu^{4}\\
 =\ & \frac{4\pi^{3}\Gamma_{md}\Omega}{\langle c^{3}\rangle}\nu^{4},
\end{align}
where \(\Omega\) is the volume per atom in the crystal unit cell, \(\langle c^{3}\rangle\) is the polarization-averaged (cubed) speed of sound in the material, B is the coefficient needed by G4CMP, and \(\Gamma_{md}\) is a mass defect coefficient capturing the average deviation of the crystal from isotopic purity \cite{TamuraIsotopeCalculationGe,TamuraIsotopicCalculationsGaAs}\footnote{The B in equation 10 is consistent with the formalism on the G4CMP GitHub respository. This B is different from equation 5.}. This expression holds true in the isotropic continuum approximation, which applies to phonons with long wavelengths relative to the interatomic spacing. The volume per atom can be calculated using
\begin{equation}
    \Omega = \frac{M}{\rho \cdot N_A},
\end{equation}
where M is the molar mass, $\rho$ is the density and $N_A$ is Avogadro's number. The mass defect coefficient \cite{Ramya,Morelli} is given by

\begin{equation}
\label{eq:MassDefectCoefficient}
\Gamma_{md}=\frac{\left\langle \overline{\Delta M^{2}}\right\rangle }{\left\langle \overline{M}\right\rangle ^{2}}.
\end{equation}

$\left\langle \overline{\Delta M^{2}}\right\rangle$ refers to the average mass variance and $\left\langle \overline{M}\right\rangle ^{2}$ represents the average mass squared.
In this development, our use of the isotropic continuum approximation implies that Equation~\ref{eq:IsotopicScattering1} represents an average scattering rate for all acoustic phonon polarizations. Although we use this definition in this work, it is worth noting that other interpretations of the mass defect coefficient have also been used \cite{Riley,Nakib}. 
By substituting these values into Equation \ref{eq:MassDefectCoefficient}, the mass defect coefficient is obtained (\(\Gamma_{md}\) = $4.65\times10^{-5}$). The volumes per atom were obtained for both \(\alpha\)-Quartz and amorphous silica to be 12.55 \AA$^3$ and 15.12 \AA$^3$.

Using the above formalism, the isotopic scattering rates are calculated for amorphous and \(\alpha\)-Quartz silica to be $1.24\cdot10^{-43}$~s$^3$ and $9.60\cdot10^{-44}$~s$^3$ respectively.
\section{Phonon Density of States (DOS)}
The phonon density of states is necessary for predicting the relative probability of exciting acoustic phonon modes. This calculation is performed for \(\alpha\)-Quartz using Quantum Espresso (QE) \cite{QE} and Phonopy \cite{Phonopy}. QE is a Python package for performing density functional perturbation theory calculations. Phonopy is a Python package designed to use results from Quantum Espresso and other DFT engines to compute phonon-specific properties. The details of this implementation are now available in a tutorial on the G4CMP Github repository \cite{StoneWhitehead2025G4CMPPhononModeDOS}.

\begin{figure}[h]
    \centering
    \includegraphics[width=1\linewidth]{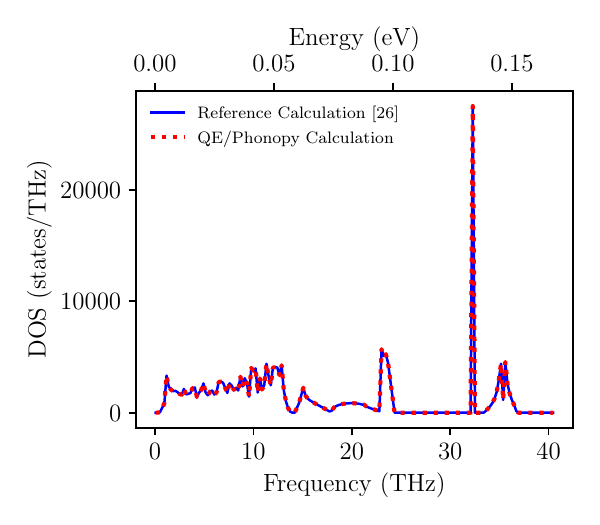}
    \caption{Total density of states (DOS) for $\alpha$-Quartz silica (blue line). (Calculation using Quantum Espresso compared to reference (red line) calculation \cite{Togo2023SiO2MDR6922})}
    \label{fig:totalDOS}
\end{figure}

\begin{figure}[h]
    \centering
    \includegraphics[width=1\linewidth]{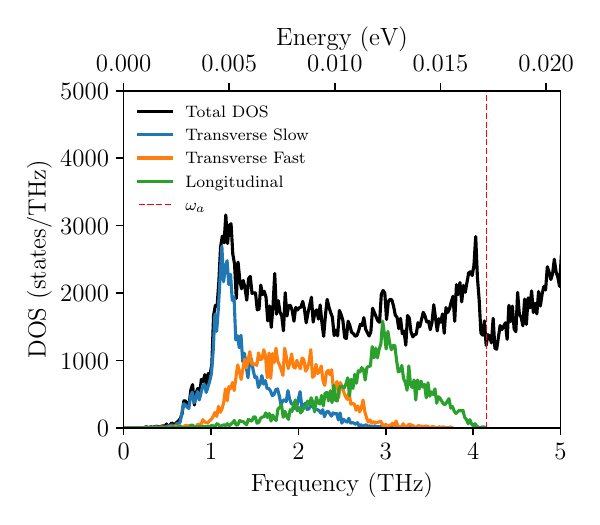}
    \caption{The phonon density of states with contributions from the first three acoustic modes: transverse slow, transverse fast and longitudinal (blue, orange and green). The dashed red line corresponds to the highest possible energy of an acoustic phonon \(\omega_{a}\).}
    \label{fig:modeDOS}
\end{figure}

To validate DOS calculations, the Materials Data Repository (MDR) \cite{Togo2023SiO2MDR6922} provides the result of many DOS and band structure calculations with reference information (Figure \ref{fig:totalDOS}). After validation of the DFPT results, the contributions from individual acoustic phonon modes are extracted (Figure \ref{fig:modeDOS}). For implementation into G4CMP, the necessary parameters are the relative DOS contributions at 1 THz (transverse slow = 0.543, transverse fast = 0.397, longitudinal
= 0.0595) and maximum acoustic phonon energy \((\omega_{a}\) = 4.2 THz). Due to the extreme computation cost required to calculate the DOS for a thermal oxide and the projected small impact on simulation results, the DOS values have been approximated as equivalent to $\alpha$-Quartz for the amorphous implementation of silica.  

\section*{Validation by Phonon Caustics}

\begin{figure}[t!]
\centering
\includegraphics[width=0.75\linewidth]{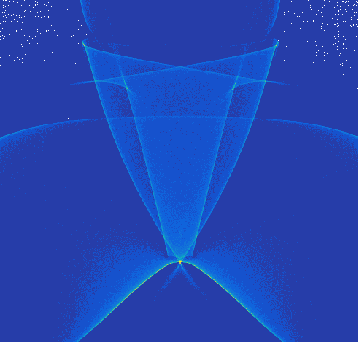}
\includegraphics[width=0.75\linewidth]{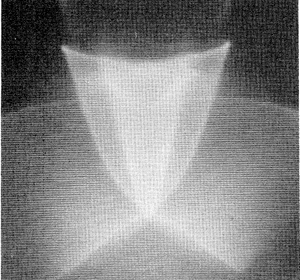}
\caption{Top: phonon caustic image for \(\alpha\)-Quartz (SiO\(_{2}\)) obtained from G4CMP. The crystal direction [1\(\bar{1}\)00] is at the center of the pattern, oriented out-of-page. Bright regions indicate directions of high phonon flux. Bottom: phonon caustic image for \(\alpha\)-Quartz measured in Ref.~\cite{KoosWolfeQuartz1984}, where the crystal direction [1\(\bar{1}\)00] is out of page and scan range is $\pm 68.19^{\circ}$.}
\label{fig:alphaSiO2CausticsSim_001}
\end{figure}

\begin{figure}[t!]
\centering
\includegraphics[width=0.78\linewidth]{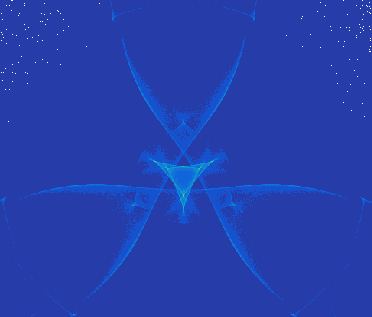}
\includegraphics[width=0.69\linewidth]{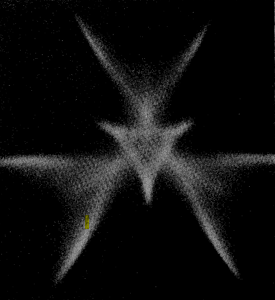}
\caption{Top: phonon caustic image for \(\alpha\)-Quartz (SiO\(_{2}\)) obtained from G4CMP. The crystal direction [0001] is at the center of the pattern, oriented out-of-page. Bright regions indicate directions of high phonon flux. Bottom: phonon caustic image for \(\alpha\)-Quartz measured in Ref.~\cite{KoosWolfeQuartz1984}, where the crystal direction [0001] is out of page and scan range is $\pm 68.19^{\circ}$.}
\label{fig:alphaSiO2CausticsSim_111}
\end{figure}

Following the standard introduced in \cite{G4CMP2025}, a validation of accurate phonon propagation for a given crystal can be performed by simulating phonon caustics. Phonon caustics refer to the intensity pattern created on the top of a sample from an isotropic point source of phonons generated at the bottom. Performing the simulation for different crystal orientations, we can then directly compare the results to corresponding bolometer images. Bolometer images are spatial images of phonon propagation across a crystal. The experimental images used for comparison (Figure \ref{fig:alphaSiO2CausticsSim_001}, Figure \ref{fig:alphaSiO2CausticsSim_111}) are measured via a superconducting Al bolometer on the top of a crystal \cite{KoosWolfeQuartz1984}. The detector measures phonon bursts from laser-induced phonon point sources on the bottom of the crystal. This allows for the reconstruction of the spatial distribution of phonons on the top of the crystal. The prebuilt example for generating phonon caustics for this validation step can be found on the G4CMP GitHub repository \cite{G4CMPCausticsExample}.

For \(\alpha\)-Quartz, our simulated caustic pattern was able to reproduce bolometer images \cite{KoosWolfeQuartz1984} of two different crystal orientations (crystal direction [1\(\bar{1}\)00] shown in Figure \ref{fig:alphaSiO2CausticsSim_001} and  [0001] shown in Figure \ref{fig:alphaSiO2CausticsSim_111}). The amorphous case demonstrates intensity only dependent on the path length from the phonon source.
\newpage
\section{Conclusions}
 The resulting amorphous implementation allows for modeling of background events in the thermal oxide layer (amorphous silica) as well as phonon propagation across this layer from events in the Si substrate. Following the analysis and validation reported in this paper, the silica materials are now available on the G4CMP github repository for use \cite{CrystalMaps2025}. Additionally, tools are provided from this work for members of the community looking to simulate new materials with G4CMP. The phonon DOS tutorial using QE and phonopy is located on the G4CMP github repository. For those repeating this analysis, the G4CMP consortium can aid in supporting this work and providing community access \cite{G4CMPConsortiumConfluence}.

\section*{Acknowledgements}
We thank Ryan Linehan and Mike Kelsey for organizing a community effort to support this work through the G4CMP consortium. We thank the rest of the G4CMP consortium for guiding this work. 

CS developed the silica modeling work presented in this paper and IH provided supporting checks. CS developed the tutorial for calculating the phonon DOS and ML revised. IH led the development of a general formalism for this work. IH derived a complete formalism for extracting Lamé parameters for a given crystal symmetry group and developed the Python implementation. CS and IH drafted the manuscript. The remaining authors provided feedback that shaped the manuscript.

This work was supported by the Gordon and Betty Moore Foundation (10.37807/GBMF11571) and the US Department of Energy, Office of Science, Office of Nuclear Physics, under award nos. DE-SC0021245 and DE-SC0023540. This manuscript has been authored by Fermi Research Alliance, LLC under Contract No.
DE-AC02-07CH11359 with the U.S. Department of Energy. This work was supported by the National Quantum Information Science Research Centers, Quantum Science Center and Illinois Institute of Technology
Department of Physics.

\clearpage
\bibliographystyle{IEEEtran}
\bibliography{main}{}

@article{Geant4,
  author    = {S. Agostinelli and others},
  title     = {{GEANT4—a simulation toolkit}},
  journal   = {Nucl. Instrum. Methods Phys. Res. A},
  volume    = {506},
  number    = {3},
  pages     = {250--303},
  year      = {2003},
  month     = jul,
}

@misc{Togo2023SiO2MDR6922,
  author       = {Atsushi Togo},
  title        = {Ab-initio phonon calculation for SiO\textsubscript{2} / P6\_222 (180) / materials id 6922},
  year         = {2023},
  publisher    = {National Institute for Materials Science},
  url          = {https://mdr.nims.go.jp/datasets/93d8ebae-5983-49de-b34c-61e055f72d0b},
  note         = {MDR dataset, published May 14, 2023}
}

@article{Allison2006,
  author       = {J. Allison and K. Amako and J. Apostolakis and H. Araujo and P. Dubois and M. Asai and G. Barrand and R. Capra and S. Chauvie and G. Depaola and J. Folger and F. Foppiano and A. Howard and H. Kimura and T. Kawabata and J. Konopka and S. Incerti and A. Johnson and V. Ivanchenko and M. Maire and P. Nieminen and T. Sasaki and D. Wright and G. Cosmo},
  title        = {Geant4 Developments and Applications},
  journal      = {IEEE Transactions on Nuclear Science},
  volume       = {53},
  number       = {1},
  pages        = {270--278},
  year         = {2006},
  doi          = {10.1109/TNS.2006.869826}
}

@article{Allison2016,
  author       = {J. Allison and K. Amako and J. Apostolakis and P. Arce and M. Asai and G. Barrand and R. Capra and S. Chauvie and R. Chytracek and G. Cosmo and P. Degtyarenko and A. Dotti and M. Dressel and F. Foppiano and A. Howard and H. Kimura and V. Ivanchenko and T. Kawabata and J. Konopka and M. Maire and P. Nieminen and T. Sasaki and D. Wright},
  title        = {Recent Developments in Geant4},
  journal      = {Nuclear Instruments and Methods in Physics Research Section A},
  volume       = {835},
  pages        = {186--225},
  year         = {2016},
  doi          = {10.1016/j.nima.2016.06.125}
}

@article{G4CMP,
  author    = {M. Kelsey and others},
  title     = {{G4CMP: A Geant4 extension for phonon and charge transport in cryogenic crystals}},
  journal   = {Nucl. Instrum. Methods Phys. Res. A},
  volume    = {1050},
  pages     = {168473},
  year      = {2023},
  doi       = {10.1016/j.nima.2023.168473},
  url       = {https://doi.org/10.1016/j.nima.2023.168473}
}

@article{G4CMP2025,
  author    = {I. Hernandez and R. Linehan and R. Khatiwada and K. Anyang and D. Baxter and G. Bratrud and E. Figueroa-Feliciano and L. Hsu and M. Kelsey and D. Temples},
  title     = {{Modeling athermal phonons in novel materials using the G4CMP simulation toolkit}},
  journal   = {Nucl. Instrum. Methods Phys. Res. A},
  volume    = {1073},
  pages     = {170172},
  year      = {2025},
  month     = apr,
  doi       = {10.1016/j.nima.2024.170172},
  url       = {https://doi.org/10.1016/j.nima.2024.170172}
}

@article{Tamura1985,
  author    = {S. Tamura},
  title     = {{Spontaneous decay rates of LA phonons in quasi-isotropic solids}},
  journal   = {Phys. Rev. B},
  volume    = {31},
  number    = {4},
  pages     = {2574--2582},
  year      = {1985},
  month     = feb,
}

@article{amorphTOEC,
  author    = {W. T. Yost and M. A. Breazeale},
  title     = {{Third-order elastic constants of fused silica}},
  journal   = {J. Appl. Phys.},
  volume    = {44},
  number    = {4},
  pages     = {1909--1913},
  year      = {1973}
}

@article{Kim2024,
  author       = {Inwook Kim and others},
  title        = {Signal processing and spectral modeling for the BeEST experiment},
  journal      = {Physical Review D},
  volume       = {111},
  number       = {5},
  pages        = {052010},
  year         = {2025},
  doi          = {10.1103/PhysRevD.111.052010},
  url          = {https://doi.org/10.1103/PhysRevD.111.052010}
}

@article{Bogardus1965,
  author    = {E. H. Bogardus},
  title     = {Third‐Order Elastic Constants of Ge, MgO, and Fused $SiO_{2}$},
  journal   = {Journal of Applied Physics},
  volume    = {36},
  number    = {8},
  pages     = {2504--2513},
  year      = {1965},
  doi       = {10.1063/1.1714520},
  url       = {https://pubs.aip.org/aip/jap/article/36/8/2504/508649/Third-Order-Elastic-Constants-of-Ge-MgO-and-Fused}
}

@article{alphaSound,
  author    = {D. L. Lakshtanov and S. V. Sinogeikin and J. D. Bass},
  title     = {{High-temperature phase transitions and elasticity of silica polymorphs}},
  journal   = {Phys. Chem. Miner.},
  volume    = {34},
  number    = {1},
  pages     = {11--22},
  year      = {2007}
}

@article{alphaTOEC,
  author    = {J. Zhao and J. M. Winey and Y. M. Gupta},
  title     = {{First-principles calculations of second- and third-order elastic constants for single crystals of arbitrary symmetry}},
  journal   = {Phys. Rev. B},
  volume    = {75},
  number    = {9},
  pages     = {094105},
  year      = {2007},
  month     = mar,
}

@article{QE,
  author    = {P. Giannozzi and others},
  title     = {{QUANTUM ESPRESSO: a modular and open-source software project for quantum simulations of materials}},
  journal   = {J. Phys. Condens. Matter},
  volume    = {21},
  number    = {39},
  pages     = {395502},
  year      = {2009}
}

@article{Phonopy,
  author    = {A. Togo and I. Tanaka},
  title     = {{First principles phonon calculations in materials science}},
  journal   = {Scr. Mater.},
  volume    = {108},
  pages     = {1--5},
  year      = {2015}
}

@article{SuperCDMs,
   title={Projected sensitivity of the SuperCDMS SNOLAB experiment},
   volume={95},
   ISSN={2470-0029},
   url={http://dx.doi.org/10.1103/PhysRevD.95.082002},
   DOI={10.1103/physrevd.95.082002},
   number={8},
   journal={Physical Review D},
   publisher={American Physical Society (APS)},
   author={Agnese, R. and others},
   year={2017},
   month=apr }

@misc{StoneWhitehead2025G4CMPPhononModeDOS,
  author       = {StoneWhitehead, Caitlyn},
  title        = {{G4CMP\_PhononModeDOS}: Phonon Mode DOS Tutorial},
  year         = {2025},
  publisher    = {GitHub},
  howpublished = {\url{https://github.com/cstonewhitehead/G4CMP_PhononModeDOS/tree/main/PhononModeDOS_Tutorial}},
  note         = {GitHub repository, accessed July 13, 2025}
}

@article{Campbell_Deem_2020_Lame,
   title={Multiphonon excitations from dark matter scattering in crystals},
   volume={101},
   ISSN={2470-0029},
   url={http://dx.doi.org/10.1103/PhysRevD.101.036006},
   DOI={10.1103/physrevd.101.036006},
   number={3},
   journal={Physical Review D},
   publisher={American Physical Society (APS)},
   author={Campbell-Deem, Brian and Cox, Peter and Knapen, Simon and Lin, Tongyan and Melia, Tom},
   year={2020},
   month=feb }

@Inbook{Fedorov1968,
author="Fedorov, Fedor I.",
title="General Theory of Elastic Waves in Crystals Based on Comparison with an Isotropic Medium",
bookTitle="Theory of Elastic Waves in Crystals",
year="1968",
publisher="Springer US",
address="Boston, MA",
pages="169--209",
isbn="978-1-4757-1275-9",
doi="10.1007/978-1-4757-1275-9_5",
url="https://doi.org/10.1007/978-1-4757-1275-9_5"
}

@book{srivastava1990physics,
  title={The Physics of Phonons},
  author={Srivastava, G.P.},
  isbn={9780852741535},
  lccn={90031835},
  url={https://books.google.com/books?id=OE-bHd2gzVgC},
  year={1990},
  publisher={Taylor \& Francis}
}

@article{TamuraIsotopicCalculationsGaAs,
  title = {Isotope scattering of large-wave-vector phonons in GaAs and InSb: Deformation-dipole and overlap-shell models},
  author = {Tamura, Shin-ichiro},
  journal = {Phys. Rev. B},
  volume = {30},
  issue = {2},
  pages = {849--854},
  numpages = {0},
  year = {1984},
  month = {Jul},
  publisher = {American Physical Society},
  doi = {10.1103/PhysRevB.30.849},
  url = {https://link.aps.org/doi/10.1103/PhysRevB.30.849}
}

@article{TamuraIsotopeCalculationGe,
  title = {Isotope scattering of dispersive phonons in Ge},
  author = {Tamura, Shin-ichiro},
  journal = {Phys. Rev. B},
  volume = {27},
  issue = {2},
  pages = {858--866},
  numpages = {0},
  year = {1983},
  month = {Jan},
  publisher = {American Physical Society},
  doi = {10.1103/PhysRevB.27.858},
  url = {https://link.aps.org/doi/10.1103/PhysRevB.27.858}
}

@article{Ramya,
title = "Alloy scattering of phonons",
author = "Ramya Gurunathan and Riley Hanus and Snyder, {G. Jeffrey}",
note = "Publisher Copyright: {\textcopyright} 2020 The Royal Society of Chemistry.",
year = "2020",
month = jun,
doi = "10.1039/c9mh01990a",
language = "English (US)",
volume = "7",
pages = "1452--1456",
journal = "Materials Horizons",
issn = "2051-6347",
publisher = "Royal Society of Chemistry",
number = "6",

}

@article{Morelli,
  title = {Estimation of the isotope effect on the lattice thermal conductivity of group IV and group III-V semiconductors},
  author = {Morelli, D. T. and Heremans, J. P. and Slack, G. A.},
  journal = {Phys. Rev. B},
  volume = {66},
  issue = {19},
  pages = {195304},
  numpages = {9},
  year = {2002},
  month = {Nov},
  publisher = {American Physical Society},
  doi = {10.1103/PhysRevB.66.195304},
  url = {https://link.aps.org/doi/10.1103/PhysRevB.66.195304}
}

@article{Riley,
    author = {Hanus, Riley and Gurunathan, Ramya and Lindsay, Lucas and Agne, Matthias T. and Shi, Jingjing and Graham, Samuel and Jeffrey Snyder, G.},
    title = "{Thermal transport in defective and disordered materials}",
    journal = {Applied Physics Reviews},
    volume = {8},
    number = {3},
    pages = {031311},
    year = {2021},
    month = {08},
    issn = {1931-9401},
    doi = {10.1063/5.0055593},
    url = {https://doi.org/10.1063/5.0055593},
    eprint = {https://pubs.aip.org/aip/apr/article-pdf/doi/10.1063/5.0055593/19741856/031311\_1\_online.pdf},
}

@article{Nakib,
  title = {Beyond the Tamura model of phonon-isotope scattering},
  author = {Protik, Nakib H. and Draxl, Claudia},
  journal = {Phys. Rev. B},
  volume = {109},
  issue = {16},
  pages = {165201},
  numpages = {10},
  year = {2024},
  month = {Apr},
  publisher = {American Physical Society},
  doi = {10.1103/PhysRevB.109.165201},
  url = {https://link.aps.org/doi/10.1103/PhysRevB.109.165201}
}

@article{Hearmon,
author = "Hearmon, R. F. S.",
title = "{`Third-order' elastic coefficients}",
journal = "Acta Crystallographica",
year = "1953",
volume = "6",
number = "4",
pages = "331--340",
month = "Apr",
doi = {10.1107/S0365110X53000909},
url = {https://doi.org/10.1107/S0365110X53000909},
}

@article{KoosWolfeQuartz1984,
  author    = {G. L. Koos and J. P. Wolfe},
  title     = {Phonon focusing in piezoelectric crystals: Quartz and lithium niobate},
  journal   = {Phys. Rev. B},
  volume    = {30},
  number    = {6},
  pages     = {3130--3140},
  year      = {1984},
  month     = sep,
  publisher = {American Physical Society},
  doi       = {10.1103/PhysRevB.30.3130},
  url       = {https://journals.aps.org/prb/abstract/10.1103/PhysRevB.30.3470}
}

@misc{G4CMPCausticsExample,
  author       = {Michael Kelsey and contributors},
  title        = {G4CMP — Caustics Example},
  year         = {2023},
  howpublished = {\url{https://github.com/kelseymh/G4CMP/tree/master/examples/caustics}},
  note         = {Accessed: 2025-05-25}
}

@misc{CrystalMaps2025,
  author       = {Caitlyn Stone-Whitehead},
  title        = {G4CMP CrystalMaps: Silica Elastic Constants and Sound Velocities},
  year         = {2025},
  howpublished = {\url{https://github.com/kelseymh/G4CMP/tree/master/CrystalMaps}},
  note         = {Includes silica-related material property data files for use with G4CMP. Accessed: 2025-05-25}
}

@misc{G4CMPConsortiumConfluence,
  author       = {{G4CMP Consortium}},
  title        = {G4CMP Consortium: Cryogenic Phonon Modeling Toolkit},
  year         = {2024},
  howpublished = {\url{https://confluence.slac.stanford.edu/x/Qk6fDQ}},
  note         = {SLAC Confluence portal for G4CMP documentation and collaboration. Accessed: 2025-05-25}
}

\end{document}